\documentclass[12pt]{article}
\usepackage{epsfig}
\usepackage{amsfonts}
\usepackage{amsopn}
\usepackage{amsmath}
\usepackage{verbatim,amsthm}

\title
{
\vskip-50 pt
\begin{flushright}
\normalsize\rm NORDITA-2012-17
\end{flushright}
\vskip 20 pt
On nonlinearity of p-brane dynamics
}

\author{
 A. A. Zheltukhin $^{a,b,c}$\thanks{e-mail: aaz@physto.se}  \\ \\
$^a$ Kharkov Institute of Physics and Technology, \\
1, Akademicheskaya St., Kharkov, 61108, Ukraine \\  
$^b$ Physics Department, Stockholm University, AlbaNova,  \\
106 91, Stockholm, Sweden \\ 
$^c$ NORDITA,  \\
Roslagstullsbacken 23, 106 91 Stockholm, Sweden
}

\date{}

\begin{document}

\maketitle

\begin{abstract}

Nonlinear equations of $p$-branes in $D=(2p+1)$-dimensional Minkowski 
space are discussed.  Presented are
new exact solutions for a set of spinning $p$-branes with 
$p=2,3,....,(D-1)/2$ and the Abelian 
symmetries  $U(1)\times U(1)\times\ldots \times U(1)$ 
of their shapes.

\end{abstract}

\section{Introduction}

Membranes and $p$-branes play an important role in M/string theory \cite{M1}, 
but the construction of general solutions for their classical 
 equations meets serious problems caused by their nonlinearity [2-13].
\nocite{tucker, hoppe1, BST, DHIS, FI, WHN, Z_0, WLN, BZ_0, hoppe2, Pol, UZ}
This preserves the necessity of the search and classification of various 
 particular solutions of the brane equations. 
Spinning membranes ($p=2$) with spherical or toroidal topology embedded 
in flat backgrounds with or without toroidal compactifications, 
as well as in $AdS_p\times S^q$ space-times, form one of the sectors 
the equations for which were considered in  \cite{CT},  \cite{KY}, \cite{HN},
\cite{hoppe6}, \cite{AFP}.
These interesting  results were  generalized to the case of
 $3$-branes ($p=3$) with more general symmetries $SU(n)\times SU(m)\times SU(k)$ by 
complexifying the target space \cite{AF} and proving the radial stability of 
these configurations. Still it is important to find exact solutions of the brane
 equations in  the explicit form, as well as to study spinning branes with any $p$. 

Here we partially touch this problem and find exact solutions for the set of closed 
spinning $U(1)^p$-invariant $p$-branes with the topology of $p$-torus embedded in 
 $D=(2p+1)$-dimensional Minkowski space. Our construction generalizes the $U(1)$
-invariant membrane anzats proposed by Hoppe in \cite{JU1} and its exact 
solutions for $D=5$ found in \cite{TZ}. 
We construct the new anzats starting from the exactly solvable one \cite{Znpb}
 and substitute $p$ propagating polar angle coordinates 
instead of $p$ propogating radial coordinates describing  $U(1)^p$-invariant 
collapsing $p$-branes.
For the case of the $U(1)^p$-invariant noncompact $p$-branes without boundaries
our anzats describes spinning branes with the topology of $p$-dimensional hyperplanes.
 The constructed exact solutions for the case $p=5$ describe $U(1)^5$-invariant 
spinning 5-branes of M/string theory in D=11 space-time.

\section{P-brane equations}

The Dirac action for a p-brane without boundaries
is defined  by the integral 
\[
S=T\int \sqrt{|G|}d^{p+1}\xi, \label{1}
\]
in the dimensionless worldvolume parameters $\xi^{\alpha}$ ($\alpha=0,\ldots,p$). 
       The components $x^{m}=(t,\vec{x})$ of the brane world vector in
 the D-dimensional Minkowski space with the signature $\eta_{mn}=(+,-,\ldots,-)$
have the dimension of length, and the dimension of tension $T$ is $L^{-(p+1)}$.
The induced metric $G_{\alpha \beta}:=\partial_{\alpha} x_{m}\partial_{\beta} x^{m}$ 
is presented in $S$ by its determinant $G$.

 After splitting the parameters $\xi^{\alpha}:=(\tau,\sigma^r)$ 
the Euler-Lagrange equations and
$(p+1)$ primary constraints generated by $S$ take the form
\begin{equation}\label{5}
\partial_{\tau}{\mathcal{P}}^{m}=-T\partial_r(\sqrt{|G|}G^{r\alpha}\partial_{\alpha}x^{m}),
 \ \ \
\mathcal{P}^{m}=T\sqrt{|G|}G^{\tau \beta}\partial_{\beta} x^{m},
\end{equation}
\begin{equation}
\tilde{T}_{r}:=\mathcal{P}^{m} \partial_{r} x_{m} \approx 0, \ \ \ \
\tilde{U}:=\mathcal{P}^{m}\mathcal{P}_{m}-T^{2}|\det G_{rs}| \approx  0, \label{6}
\end{equation}
where $\mathcal{P}^{m}$ is the energy-momentum density of the brane.

It is convenient to use the orthogonal gauge simplifying the metric $G_{\alpha\beta}$  
\begin{eqnarray} \label{7}
L\tau=x^0\equiv t, \ \ \ \ G_{\tau r}= -L(\dot{\vec{x}} \cdot \partial_r \vec{x})=0, \\
 g_{rs}:=\partial_r \vec{x} \cdot \partial_s \vec{x}, \ \ \ \
G_{\alpha\beta}=\left( \begin{array}{cc}
                       L^{2}(1- {\dot{\vec{x}}}^2)& 0    \\
                         0 & -g_{rs}
                              \end{array} \right)
                               \nonumber
\end{eqnarray}
with $\dot{\vec{x}}:=\partial_{t}\vec{x}= L^{-1}\partial_{\tau}\vec{x}$.
The solution of the constraint $\tilde{U}$ (\ref{6})
takes the form
\begin{equation}\label{9}
\mathcal{P}_0=\sqrt{\vec{\mathcal{P}}^2+T^{2}|g|}, \ \ \ \
g=\det(g_{rs})
\end{equation}
and becomes the  Hamiltonian density $\mathcal{H}_0$ of the p-brane
since $\dot{\mathcal{P}}_0=0$ in view of Eq. (\ref{5}).
Using the definition of $\mathcal{P}_0$  (\ref{5}) and
 $G^{\tau\tau}={1}/{ L^{2}(1-\dot{\vec{x}}^2)}$ 
 we find the  expression of $\mathcal{P}_0$ as a function of the p-brane  velocity $\dot{\vec{x}}$
\begin{equation}\label{5'}
\mathcal{P}_0:=TL\sqrt{|detG|}G^{\tau\tau}=T\sqrt{\frac{|g|}{1-\dot{\vec{x}}^2}} \,\,\,.
\end{equation}
Taking into account this expression for $\mathcal{P}_0$ and the definition (\ref{5}) one 
can present $\vec{\mathcal{P}}$ and its evolution equation (\ref{5}) as
\begin{equation}\label{13}
\vec{\mathcal{P}}=\mathcal{P}_0 \dot{\vec{x}}
 , \ \ \ \
\dot{\vec{\mathcal{P}}}= T^{2}\partial_r \left( \frac{|g|}{\mathcal{P}_0} g^{rs}\partial_s 
\vec{x}\right).
\end{equation}
Then Eqs. (\ref{13}) yield the second-order PDE for $\vec{x}$
\begin{equation}\label{xeqv}
 \ddot{\vec{x}}=\frac{T}{\mathcal{P}_0}\partial_r \left( \frac{T}{\mathcal{P}_0}|g|g^{rs}
\partial_s \vec{x}\right).
\end{equation}
These equations may be presented in the canonical Hamiltonian  form
\[
\dot{\vec{x}}=\{H_{0}
,\vec{x}\}, \ \ \ \ \dot{\vec{\mathcal{P}}}=\{H_{0},\vec{\mathcal{P}}\}, \ \ \ \
\{\mathcal{P}_i(\sigma), x_j(\tilde{\sigma})  \}=
\delta_{ij}\delta^{(p)}(\sigma^r-\tilde{\sigma}^r)
\]
using the integrated Hamiltonian density (\ref{9}) $\mathcal{H}_0(=\mathcal{P}_0)$
\begin{eqnarray}\label{hampbr}
H_{0}=  \int d^p \sigma \sqrt{\vec{\mathcal{P}}^2+T^{2}|g|}.
\end{eqnarray}
The presence of square root in (\ref{hampbr}) points to the presence of the known 
residual symmetry preserving the orthonormal gauge (\ref{7}) 
\begin{equation}\label{diff}
\tilde{t}=t, \ \ \ \ \tilde{\sigma}^r=f^r(\sigma^s)
\end{equation}
and generated by the  constraints $\tilde{T}_r$  (\ref{6}) reduced to the form
\begin{equation}\label{T}
T_r:=\vec{\mathcal{P}}\partial_r\vec{x}=0 \ \ \    \Leftrightarrow \ \ \ 
 \dot{\vec{x}} \partial_r\vec{x}=0,\ \ \ (r=1,2, \ldots, p).
\end{equation}
 The freedom allows to impose
 $p$ additional time-independent conditions on $\vec{x}$ and 
 its space-like derivatives. The above description is valid for any 
 space-time and brane worldvolume dimensions $(D,p)$ with $p<D$.

\section {$\mathbf{U(1)\times U(1)\times\ldots \times U(1)}$ spinning $p$-branes
}

Here we continue studying  p-branes which evolve in $D=(2p+1)$-dimensional 
Minkowski space-time starting from the general representation for 
the $2p$-dimensional $p$-brane Euclidean vector $\vec{x}$ by $p$ pairs of 
its "polar" coordinates 
\begin{eqnarray}\label{Gen} 
\vec{x}^T(t,\sigma^{r})=
(q_1\cos\theta_1,q_1\sin\theta_1,\ldots,q_p\cos\theta_p,q_p\sin\theta_p).
\end{eqnarray}

The polar brane coordinates  $q_a=q_a(t,\sigma^{r})$ with $a = 1,\ldots,p$ and 
$\theta_a=\theta_a(t,\sigma^{r})$ certainly  depend on 
all the parameters $(t,\sigma^{1},..., \sigma^{p})$ of the p-brane 
worldvolume. However, such a dependence obstructs exact solvability of the 
brane equations (\ref{xeqv}).
 The separation of the time and $\sigma^{r}$ variables, 
fixed by the choice
 $q_a=q_a(t)$ and $\theta_a=\theta_a(\sigma^{r})$, transforms
 the representation (\ref{Gen}) to 
an exactly solvable anzats describing contracted 
toroidal $p$-branes \cite{Znpb}.

Here we consider an alternative anzats originating in (\ref{Gen}), but 
with the  "polar" angles $\theta_a=\theta_a(t)$ independent 
of  $\sigma^{r}$ and the time-dependent "radial" coordinates $q_a=q_a(\sigma^r)$
 \begin{eqnarray} 
\vec{x}^T
=(q_1\cos\theta_1,q_1\sin\theta_1,q_2\cos\theta_2,q_2\sin\theta_2,
\ldots,q_p\cos\theta_p,q_p\sin\theta_p), \label{anzats}
 \\
q_a=q_a(\sigma^{r}),\,\,\, 
 \theta_a=\theta_a(t). 
\, \, \, \, \,  \, \, \,\, \, \, \, \,  \, \, \,\, \, \, \, \,  \, \, \,\, \, \, \, \,  \, \, \,\, \, \, \, \,  \, \, \, \, \, \, \, \,  \, \, \,\nonumber
\end{eqnarray}
We show that anzats (\ref{anzats}) is  also exactly solvable
and describes either spinning toroidal  $p$-branes
or spinning $p$-dimensional hyperplanes.

The brane space vector (\ref{anzats}) lies in the $2p$-dimensional 
Euclidean subspace of the 
 $(2p+1)$-dimensional Minkowski space and automatically satisfies to 
 the orthogonality 
 constraints (\ref{T}): $\dot{\vec{x}} \partial_r\vec{x}=0$.
 At any fixed moment of time  the  world vector $\vec{x}^T$
(\ref{anzats}) is produced  from 
 $\vec{x_0}^T=(q_1,0,q_2,0,\ldots,q_p,0 )$
by rotations of the diagonal subgroup 
$U(1)^p\in SO(2p)$,
  parametrized by the angles $\theta_{a}(t)$ describing  rotations in the 
planes $x_1x_2$, $x_3x_4$ ,...,$x_{2p-1}x_{2p}$.
Each of these $U(1)$ symmetries is 
locally isomorphic to one of the  $O(2)$ subgroups of the $SO(2p)$ 
group of the Euclidean rotations.   
 Thus, the p-dimensional hypersurface  $\Sigma_{p}$ of $U(1)^p$-invariant
 p-brane  has the Abelian group 
$ U(1)\times U(1)\times\ldots \times U(1)\equiv U(1)^p$ 
as its isometry containing $p$ Killing vectors.
The Abelian character of the group $U(1)^p$ supposes the existence 
of a local parametrization of the
 p-brane hypersurface $\Sigma_{p} $ with the metric tensor $g_{rs}$
 independent of $\sigma^r$.
Actually, the metric $G_{\alpha \beta}$ 
 of the $(p+1)$-dimensional worldvolume $\Sigma_{p+1}$
created by (\ref{anzats}) has the form  similar to (\ref{7}) with 
the non-zero components 
\begin{eqnarray}\label{metrik} 
 G_{tt}=1- \sum_{a=1}^{p}(q_{a}\dot\theta_{a})^2, \  \  \  
g_{rs}= \sum_{a=1}^{p}q_{a,r}q_{a,s}\equiv \mathbf{q}_{,r}\mathbf{q}_{,s},  
\  \  \  
\mathbf{q}:=(q_1,..,q_p),
\end{eqnarray}  
where  $\dot{\theta_{a}}\equiv \partial_{t}\theta_{a}$,  
\ \  $q_{a,r}\equiv \partial_{r}q_{a}$
and yields the following interval $ds^2$ on $\Sigma_{p+1}$ 
\begin{equation}\label{intr}   
ds^2_{p+1}=
(1- \sum_{a=1}^{p}(q_{a}\dot\theta_{a})^{2}) dt^{2} - \sum_{a=1}^{p}dq_{a}dq_{a}.
\end{equation}
 This representation shows that the change of $\sigma^{r}$ parametrizing
 $\Sigma_{p}$  by the new local coordinates $q_{a}(\sigma^{r})$
 makes the induced metric on $\Sigma_{p}$  independent 
of  $\sigma^{r}$.  
 In the new parametrization the above mentioned 
 Killing vector fields on $\Sigma_{p}$ take the form of the 
 derivatives  $\frac{\partial}{\partial q_a}$.

This shows that the closed hypersurface $\Sigma_{p}$ presented   by 
(\ref{anzats}) is a flat  manifold 
with its shape isomorphic to a flat p-dimensional 
torus $S^1\times S^1\times \ldots \times S^1$ 
with p multipliers $S^1$.
The  next step is to show that anzats (\ref{anzats}) is in fact an exact 
solution of the nonlinear chain  (\ref{xeqv}).

\section{Solutions of spinning brane equations 
}

The substitution of anzats (\ref{anzats}) into Eqs.(\ref{xeqv})
 reduces these $2p$ nonlinear PDEs for the $\vec x$ components 
to $p$ equations for the $p$ components of $\mathbf{q}(\sigma^{r})$ 
\begin{equation}\label{rxeqv'}
\dot\theta_{a}^{2}q_{a} + 
\frac{T}{\mathcal{P}_0}\partial_r \left( \frac{T}{\mathcal{P}_0}|g|g^{rs}
\partial_s q_{a}\right)=0.
\end{equation}
Using the relation $\dot{\vec{x}}^2= \sum_{a=1}^{p}q_{a}^{2}\dot\theta_{a}^{2}$  
 one can present the energy density   $\mathcal{P}_0$ (\ref{5'}) 
as a function of the velocity components $\dot{\theta_{a}}(t)$
\begin{equation}\label{5''}
\mathcal{P}_0=T\sqrt{\frac{|g|}{1-\sum_{a=1}^{p}q_{a}^{2}\dot\theta_{a}^{2}}}.
\end{equation}
Taking into account the energy density conservation $\dot{\mathcal{P}}_0=0$ 
together with the time-independence 
of the metric $g_{rs}= \mathbf{q}_{,r}\mathbf{q}_{,s}$ (\ref{metrik})
we obtain the solution for the polar angles $\theta_{a}(t)$
\begin{equation}\label{solteta}
\theta_{a}(t) =\theta_{0a} + \omega_{a}t,
\end{equation}
where $\theta_{0a}$ and  $\omega_{a}$ are the integration constants.  
Then  $\mathcal{P}_0$ (\ref{5''}) asquires the form
\begin{equation}\label{5'''}
\mathcal{P}_0(\sigma^r)=T\sqrt{\frac{|g|}{1-\sum_{a=1}^{p}\omega_{a}^{2}q_{a}^{2}}} \,  
\end{equation}
which  shows  
that the coordinates $\theta_{a}$ are cyclic and their conjugate momenta
$j_a:=\frac{\partial \mathcal{L}}{\partial \dot{\theta_a}}=
\vec{\mathcal{P}}\frac{\partial{\dot{\vec{x}}}}{\partial \dot{\theta_a}}$
are the integrals of motion  
\begin{equation}\label{tetimp} 
\frac{d j_a}{dt}=0, \ \ \
 j_a=\mathcal{P}_0 q_{a}^2\dot{\theta_a}
\equiv\mathcal{P}_0 \omega_{a}q_{a}^2,
   \ \ (a=1,2,.., p)
\end{equation}
additional to the Hamiltonian density $\mathcal{H}_0$ (\ref{9}) taking the form 
\begin{equation}\label{rotahamd}
\mathcal{H}_0=\sqrt{\sum_{a=1}^{p}(j_{a}/q_{a})^2 + T^{2}|g|}. 
\end{equation}

The functions $j_a$ have the physical sense 
of the components of the angular momentum density associated with 
the generators of rotations in the 
planes $x_1x_2$, $x_3x_4$ ,...,$x_{2p-1}x_{2p}$ 
which form the above discussed abelian group  $U(1)^p$
 \begin{equation}\label{angmom} 
j_b=
T\omega_{b}q_{b}^2 \sqrt{\frac{|g|}{1-\sum_{a=1}^{p}\omega_{a}^2q_{a}^{2}}}.
\end{equation}

To solve Eqs. (\ref{rxeqv'}) we fix the gauge  
for the residual gauge symmetry (\ref{diff}) 
\begin{equation}\label{rgaug}
q_{1}(\sigma^{r})=q_{1}(\sigma^{1}),\ \ \  q_{2}(\sigma^{r})=q_{2}(\sigma^{2}),\ ....,\ 
q_{p}(\sigma^{r})=q_{p}(\sigma^{p})
\end{equation}
that  diagonalizes the derivatives $q_{a,s}=\delta_{as}\acute{q}_{s}$, 
where $\acute{q}_{s}:=\frac{dq_s}{d\sigma^s}$, and $g_{rs}$ 
\begin{equation}\label{metr}
g_{rs}=\delta_{rs}\acute{q}_{s}^{2}, \ \ 
 g^{rs}=\frac{\delta_{rs}}{\acute{q}_{s}^{2}}, \ \ 
g= \prod_{a=1}^{p}\acute{q}_{a}^{2}\equiv \prod\acute{q}_{a}^{2}.
\end{equation}
Using the representation  
 $g g^{rs}=\frac{\delta_{rs}}{\acute{q}_{r}^{2}}\prod\acute{q}_{a}^{2}$
one can present  Eqs. (\ref{rxeqv'}) as 
\begin{equation}\label{rxeqvg}
\omega_{r}^{2}q_{r} + \frac{1}{2\acute{q}_{r}} 
\partial_{r}\left(\frac{T}{\mathcal{P}_0}\right)^2
\prod\acute{q}_{a}^{2}
+\left( \frac{T}{\mathcal{P}_0}\right)^2
\frac{q''_{r}}{\acute{q}_r^2} \prod\acute{q}_{a}^{2}=0.
\end{equation}
Then taking into account the representation
\begin{equation}\label{rels}
\left( \frac{T}{\mathcal{P}_0}\right)^2
=\frac{1-\sum_{a=1}^{p}\omega_{a}^2q_{a}^{2}}{\prod\acute{q}_{a}^{2}}
\end{equation}
and calculating its  partial derivatives in $\sigma^r$
\begin{equation}\label{derels}
\frac{1}{2}\partial_{r}\left( \frac{T}{\mathcal{P}_0}\right)^2
=- \omega_{r}^{2}q_{r}\frac{\acute{q}_{r}}{\prod\acute{q}_{a}^{2}}
-\left( \frac{T}{\mathcal{P}_0}\right)^2
\frac{q''_{r}}{\acute{q}_r}
\end{equation}
we observe that the additives composing the second term in (\ref{rxeqvg}) 
 are mutually cancelled with its first and third terms. 
This proves that anzats (\ref{anzats})  written down in the gauge (\ref{rgaug}),
 containing $p$ arbitrary periodic functions of one variable $q_a(\sigma^{a})=q_a(\sigma^{a}+2\pi)$,
\begin{eqnarray}\label{ganzats} 
\vec{x}^T
=(q_1(\sigma^{1})\cos\omega_{1}t,q_1(\sigma^{1})\sin\omega_{1}t,
\ldots,
q_p(\sigma^{p})\cos\omega_{p}t,q_p(\sigma^{p})\sin\omega_{p}t)
\end{eqnarray}
 is an exact solution of the 
spinning $p$-brane equations with the fixed initial 
data $\theta_{0a}=0$ at $t=0$. This solution describes flat toroidal $p$-branes. 

The periodicity  conditions for $q_{a}$
do not permit to consider $p$-dimensional 
hyperplanes as alternative  flat $p$-brane hypersurfaces. 
However, one can consider $p$-branes with boundaries 
and to suppose that $\sigma^{r}\in [0,\infty)$.
 Then it is possible to  choose the gauge (\ref{rgaug}) with the linear 
functions $q_{s}(\sigma^s)$
\begin{equation}\label{hpgaug}
q_{1}(\sigma^{r})=l_{1}+ k_{1}\sigma^{1},\ \ \ 
 q_{2}(\sigma^{r})=l_{2}+k_{2} \sigma^{2},\ .... \ ,
q_{p}(\sigma^{r})=l_{p}+ k_{p}\sigma^{p},
\end{equation}
where $l_a, \,  k_{a}$ are arbitrary constants with the dimension of length.

As a result, the anzats (\ref{ganzats}) added by (\ref{hpgaug}) turns out 
to describe spinning $p$-dimensional hyperplanes embedded in 
the $2p$-dimensional Euclidean space.

\section{Summary}

New exact solutions describing $U(1)^p$-invariant spinning $p$-branes embedded in 
the $D=(2p+1)$-dimensional Minkowski space are constructed.  
Solutions for the compact hypersurfaces $\Sigma_{p}$ of closed 
 $p$-branes (with $p=2,3,...,(D-1)/2$ ) are shown to be isometric to 
 flat $p$-dimensional tori with time-independent radii.
The solutions for the noncompact hypersurfaces of $U(1)^p$-invariant 
$p$-branes without boundaries are found to be presented 
by ${p}$-dimensional hyperplanes. 
The Hamiltonians desribing these spinning ${p}$-branes are presented. 
 
The found exact solutions, in particular, for $p=5$ present another exact solution 
for the 5-brane of $D=11$ M/string theory additional to the solution 
describing the $p$-brane collapse \cite{Znpb}. 
The compact solutions may be generalized, in particular, to the Minkowski spaces 
with toroidal compactifications.

\noindent{\bf Acknowledgments}

The author is grateful to M. Axenides and E.G. Floratos for their useful 
comments, the references \cite{AFP},\cite{AF} and 
to Physics Department of Stockholm University,
Nordic Inst. for Theor. Physics Nordita and ITP of Wroclaw University
for kind hospitality. 
This research was supported in part by Nordita.

\end{document}